\documentstyle[prc,aps,psfig]{revtex}
\newcommand{\be}{\begin{eqnarray}}
\newcommand{\ee}{\end{eqnarray}}

\newcommand{\bi}{\begin{itemize}}
\newcommand{\ei}{\end{itemize}}
\begin{document}
\twocolumn[\hsize\textwidth\columnwidth\hsize
           \csname @twocolumnfalse\endcsname
\title{Squeezing mode in nuclear collisions }
\author{Klaus Morawetz$^1$ and Pavel Lipavsk\'y$^2$}
\address{$^1$LPC-ISMRA, Bld Marechal Juin, 14050 Caen
 and  GANIL,
Bld Becquerel, 14076 Caen Cedex 5, France\\
Max-Planck-Institute for the Physics of Complex Systems, 
Noethnitzer Str. 38, 01187 Dresden, Germany\\
$^2$ Institute of Physics, Academy of Sciences, Cukrovarnick\'a 10, 
16200 Praha 6, Czech Republic}
\maketitle
\begin{abstract}
The time dependent Schroedinger equation
is solved analytically for a simplified model of moving infinite walls.
A new knock-out mode is described which might occur during heavy ion
collisions. The outer shell-nucleons are ionised due to the increase of
level energy when two nuclei are
approaching fast enough. This is
analogous to the Mott effect but in contrast occurs only if the
reaction time is short enough that no common ionisation threshold 
in the compound system is established. To demonstrate this pure 
nonequilibrium effect a simulation of realistic heavy ion collision by
a nonlocal Boltzmann equation is performed.

\end{abstract}
\vskip2pc]

When a system with bound states is exposed to a compression beyond
certain values the bound states break off and decompose into their 
constituents. This pressure ionisation known as the Mott transition is 
well established
in different fields of physics\cite{kker86,SRS90,SSAR95,BDGP98}. 
Alternatively in a many body system
the density could be increased. The theoretical treatments agree in
that the ionisation threshold is lowered faster than the binding
energy with increasing density which leads to a cross-over and
ionisation at the Mott density. 
These treatments rely on the fact that
one has a certain degree of equilibration in the system, at least
that the system has one unique ionisation threshold. 

This situation is
however somewhat different when processes occur far from equilibrium. 
Then there might be not enough time to establish a common threshold in the
system. In particular if two nuclei approach each other in a heavy
ion collision it takes a certain time before a compound system is
established or  decomposition happens at higher energies. One can
easily imagine that there will be no common ionisation threshold for
the 2 nuclei at the early stage of reaction. Instead we will show that
this leads to a new escaping mode by squeezing states which should
lead to the nonequilibrium emission of  particles. The principle 
phenomena has already 
been investigated in the past as light nonequilibrium particle
emission \cite{Ca88}. Here we will show that with the help of an
exactly solvable model of time dependent Schroedinger equation a new
effect arise not described so far. The different features are
transversal angular distribution and a lower bound of projectile
energy that this squeezing mode to happen. There are
experimental signals for dynamical particle emission \cite{T96,Y99}.
In contrast to diabatic emission of particles \cite{GN86} limited to
beam energies much below the Fermi energy we consider
here the case of faster processes around the Fermi energy. 
After presenting the exactly solvable model we will confirm this mode
by realistic simulations which will show a transversal
distribution and low energy of emitted particles while 
diabatic emitted particles are longitudinal peaked. 

Another intuitive picture is the following. The Pauli principle will forbid
over-occupation of states, which should result in a Fermi gap which is closed
during dissipation. Before this quasi - equilibration happens one has
essentially the situation that an outer nucleon feels a rapidly
increasing force due to the other nuclei and the Pauli - forbidden areas in
phase space. If the speed of nuclei is high there is no many - body
equilibration or dissipation but rather a shrinking of phase space for
the outer nucleons during the time after first touching of the two
nuclei. Therefore it is reasonable to assume a picture where the
boundary of the outer nucleons is shrinking with time. We will solve
such a time dependent Schroedinger equation to show that indeed the
level energy of the outer nucleons goes up and eventually leads to
ionisation. Since this is opposite to the  Mott transition
described above where the threshold decreases we call this mode squeezing mode in
the following. There are similarities to the elevator resonant
activation mode \cite{SJ93} where a time dependent potential
inside the wall creates resonant levels which trap the particles and
lift them above the barrier. Here we merely squeeze the wall.

Let us assume that the outer shell nucleons are bound states which can
be parameterised by a simple one-dimensional infinite wall model, 
i.e. a free particle  inside an infinite wall of distance $b$ at
the initial time $t=0$. Then the binding energy is
[$k_n={\pi n\over b}$]
\be
E_{n}^0={\hbar ^2 k_n^2\over 2 m}
\label{eo}
\ee
and the normalised wave function 
\be
\Psi(0,x)=\sqrt{{2\over b}} \sin{(k_n x)} {\rm e}^{i \phi}
\label{init}
\ee 
where we note that the physical state is undetermined up to a phase
$\phi$ which will be employed to find a solution of the time
dependent Schroedinger equation.
We now solve the time dependent Schroedinger equation with the
boundary condition that the wall is moving inwards with the speed
$v$. The mathematical problem is
\be
(i\hbar {\partial \over \partial t} + {\hbar^2\over 2 m}{\partial^2 \over
  \partial x^2})\Psi(t,x)&=&0\nonumber\\
\Psi(t,0)=\Psi(t,b-v t)&=&0
\label{prob}
\ee
together with the initial state (\ref{init}). Of course the moving wall can be
equivalently formulated as a {\it time dependent} $\delta$- or
Hill-Wheeler potential. This case can be considered as a special
case of \cite{DKN93}. The normalised solution reads
\be
\Psi(t,x)=\sqrt{{2\over b-v t}} \sin{{k_n x b\over b-v t}} \exp{\!\!\left
    (
\!\!-i
  {m^2 v x^2 \!\!+\!\!k_n^2 \hbar^2 b t\over 2 m \hbar (b-v t)}\right )}
\label{sol}
\ee 
which determines the phase $\phi$, especially it leads to the initial
phase  at $t=0$
\be
\phi=-{x^2 m v \over 2 b \hbar}.
\ee
Insertion into (\ref{prob}) verifies the solution. In \cite{DKN93}
such classes of solutions have been used to expand any initial condition
at $t=0$. Here we want to point out that already the basic solution
(\ref{sol}) with the extra phase $\phi$ at inital time $t=0$ bears a
physical meaning in itself.    
The solution (\ref{sol}) represents a non-separable solution. Other
classes of potentials which admit a separable solution can be found
in \cite{ES94}. 

The probability density $\rho$ and current $s$ are easily computed
\be
\rho(x,t)&=&|\Psi|^2={2\over b-v t} \sin^2{{\pi n x \over b-v t}}\nonumber\\
s(x,t)&=&2\, {\rm Im} \Psi^+ \partial_x \Psi=\rho(x,t) {x v\over vt-b}
\ee
with $\dot \rho+\partial_x s=0$ as it should.
The kinetic energy becomes
\be
E_n(t)&=&-\int dx\Psi {\hbar ^2\nabla^2\over 2 m }\Psi^+\nonumber\\
&=&{m v^2 \over 12 n^2 \pi^2} (2 n^2 \pi^2 -3 ) + {n^2 \pi^2\hbar^2\over
2 m  (b-v t)^2}
\ee
which shows that even at time $t=0$ the initial binding energy
(\ref{eo}), $E_n^0={\hbar^2 \pi^2 n^2\over 2 m b^2}$, is shifted by
\be
E_\phi={E_{\rm proj} \over 3} \left (1-{3 \over 2 n^2 \pi^2}\right )
\ee
with the projectile energy $E_{\rm proj}=m v^2/2$ due to the finite
velocity $v$ corresponding to the finite phase $\phi$.
This means that due to a finite velocity $v$ of the projectile, the
target system gets a phase jump $\phi$ and an energy $E_\phi$
immediately when they touch each other. 
The kinetic energy of a level increases with time as 
\be
E_n(t)=E_\phi+E_n^0 {b^2\over (b-v t)^2}.
\ee
Ionisation happens when this energy becomes larger or equal a now
introduced threshold $E_c$ such that the effective initial binding energy
would be $E_n^b=E_c-E_n^0$. Consequently we obtain for ionisation
\be
{b^2\over (b-v t)^2} \ge {E_c - E_\phi \over E_n^0}.
\ee
Therefore we find immediate ionisation if the phase kinetic energy $E_\phi$
exceeds the threshold $E_c$
\be
E_{\rm proj}\ge {3\over 1-{3 \over 2 n^2 \pi^2}}E_c.
\label{proj}
\ee
For projectile kinetic energies below we have to wait long enough to reach the 
threshold which leads to
\be
{b\over v}\ge t\ge {b\over v} \left (
1-
\sqrt{E_n^0 \over E_c-E_\phi}
\right )
\label{slow}
\ee
where the first inequality comes from the restriction of the model in
that there should be space between the walls left at all.

Therefore we have two cases.
If the projectile is fast enough (\ref{proj}), roughly larger than
Fermi energy, we knock out particles at the first instant of
touching. This can be seen in analogy to the observation of
\cite{Ca87} where a model of instantly removed walls was studied. Due to
the time dependent solution here we can give the velocity criterion
where such effect should happen. 

For slower projectiles and fast enough reaction
time to prevent many - body
equilibration and larger reaction time than the critical 
time (\ref{slow}) we will have a
knock-out of the outer shell nuclei as well. Both cases should be
considered as a nonequilibrium mode. 

The latter condition for projectiles below or around Fermi energy,
(\ref{slow}),  can
be translated into a geometrical condition. For the model we assume
two  approaching spherical nuclei with equal radii $R$. The case of
different radii is straightforward. The impact parameter $B$ should then be smaller than
the sum of radii of the two nuclei
$B<2 R$
in order to allow the necessary overlap. Assuming constant projectile
velocity $v_p$, the time between the first touching of the nuclei and 
the closest approach is
\be
t_m={\sqrt{(2 R)^2-B^2}\over v_p}\approx {R-B/2\over v}.
\label{tm}
\ee
Here we approximate the relative
velocity $v$ of the nuclei by a constant velocity such that at a time
$t=0$ we have
the distance $R$ and at the time of maximal overlap we have the
distance $B/2$, and the effective velocity 
$v=v_p(R-B/2)/\sqrt{(2 R)^2-B^2}$ which leads to (\ref{tm}).

The condition for reaction time (\ref{slow}) together with $2 R>B$
translates now into a
condition
for the geometry
\be
B\le 2 R \sqrt{E_n^0 \over E_c-E_\phi} \le 2R  
\label{cond}.
\ee

In other words the condition (\ref{cond}) gives the simple restriction
on the reaction geometry concerning radii and impact parameter for
which an outer bound state characterised by the binding energy $E_n^0$
is ionised. Rewriting (\ref{cond}) the condition for the
projectile velocity reads
\be
&&{3\over 1-{3 \over 2 n^2 \pi^2}} E_c\ge E_{\rm proj}
%\nonumber\\&&\qquad\qquad 
\ge {3\over 1-{3 \over 2 n^2 \pi^2}} \left (E_c-E_n^0 \left ({2 R\over
    B}\right )^2 \right ).
\nonumber\\
&&\label{proj1}
\ee
Therefore we have two projectile energy ranges. Either (\ref{proj})
for which ionisation happens immediately or 
for lower projectile energies than (\ref{proj1}) we have
overlap and possible ionisation if the projectile energy is not too
low (\ref{proj1}). These lower boundary in projectile energy clearly
distinguishes this model from the diabatic model \cite{GN86}.

\begin{figure}
\psfig{file=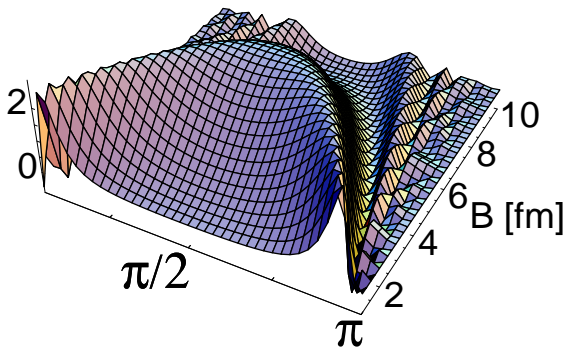,width=8cm,angle=0}
\caption{The Wigner function (\protect\ref{Wig}) integrated over the
  spatial coordinates and zero momentum versus angle and impact
  parameter
according to $b-vt=B/\sin{(\alpha)}$. The projectile energy is around
Fermi energy. For nonzero momenta the
transversal maximum is shifted to lower impact parameter.\label{rho}}
\end{figure}

In order to draw possible predictions to observable effects one would
like to have the characteristic angular distribution of such emitted
particles due to phase space squeezing. Since we do not have the exact
solution of the two dimensional geometry we associate approximatively
the emission angle with the ratio of impact parameter to the distance
of nuclei $\sin{\alpha}=B/(b-v t)$ which corresponds to the above
geometrical model. The angular distribution will be according to the
momentum of the Wigner function. The latter one can be given with
(\ref{sol})
\be
f(p,R,t)&=&\int\limits_{-(b-vt)}^{b-v t} d r {\rm e}^{- i r p}
\Psi^+(R-\frac r 2,t)\Psi(R+\frac r 2,t)
\nonumber\\
&=&2 {\sin{\xi}\over \xi} \left ( {\xi^2 \cos{(n \pi)} \over \xi^2-n^2\pi^2}
-\cos{({2\pi n R\over b-v t})} \right )
\label{Wig}
\ee 
and $\hbar \xi=p(b-v t)+m v R$. Integrating over all allowed spatial
coordinates and using the angular relation as described above we
obtain the angular distribution which is momentum and impact parameter
dependent. In figure \ref{rho} we see that for small impact parameter
indeed there is a longitudinal emission pattern in agreement with the
finding of the diabatic model \cite{CN84,Ca88}. But for specific impact
parameter around $6-8$fm for a projectile energy around the Fermi
energy we see a clear transversal distribution.

When such ionisation due to phase space squeezing happens, one can 
expect small
energies beyond the threshold. Therefore it should be possible to
observe such a mode as emitted particles with
small total energy and the angular distribution should be
transversal symmetric for the emitted particles. A good plot to 
observe this mode would be the
transverse energy versus the ratio of the transverse to total
energy. The usual multifragmentation products are on a clear
correlation line between these two observables. The described mode here
should be visible as a correlation with small transverse energy but
a large ratio between transverse and total energy. 

\begin{figure}
\parbox[]{20cm}{
\psfig{file=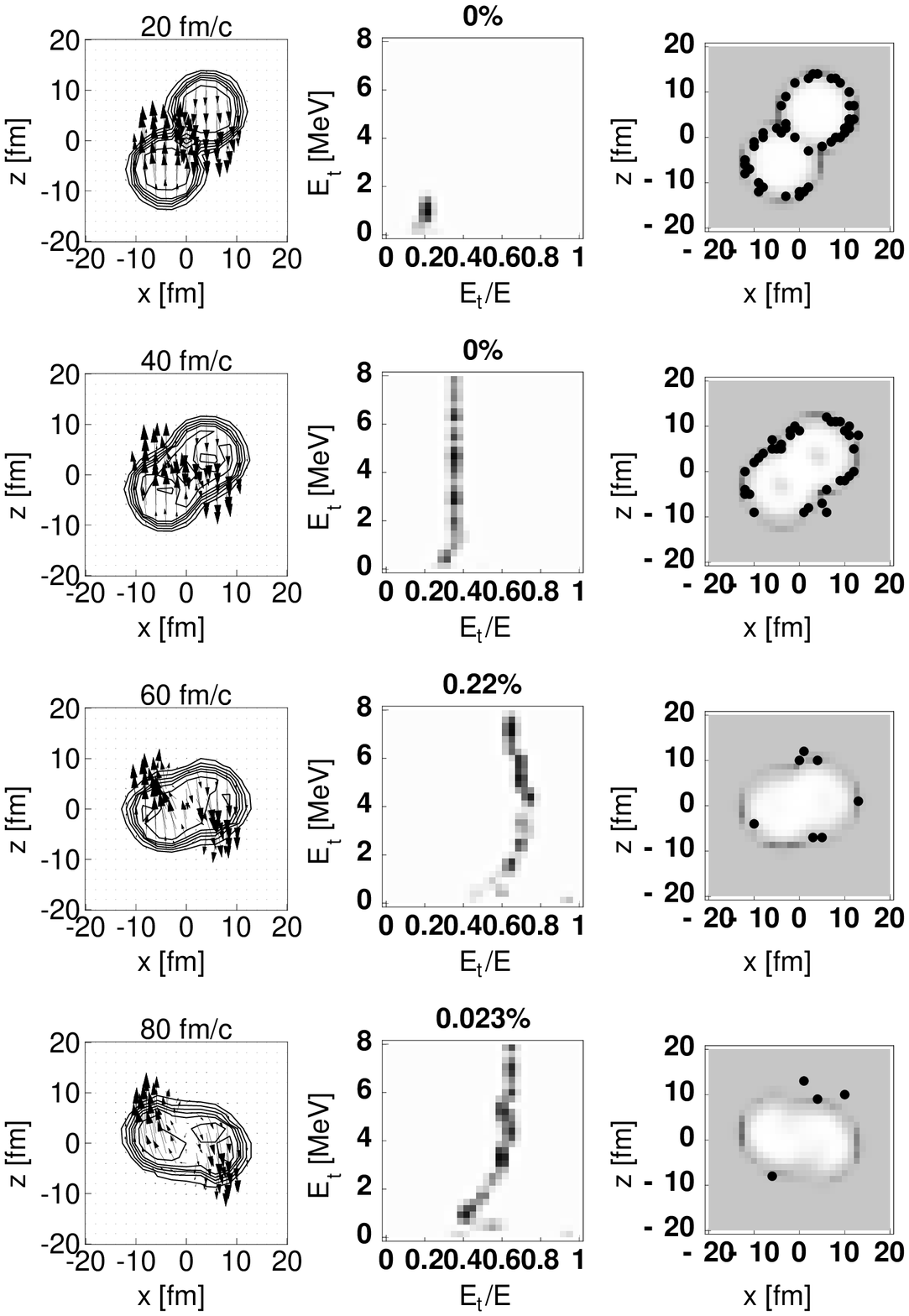,width=15cm,angle=0}}

\parbox[]{20cm}{
\psfig{file=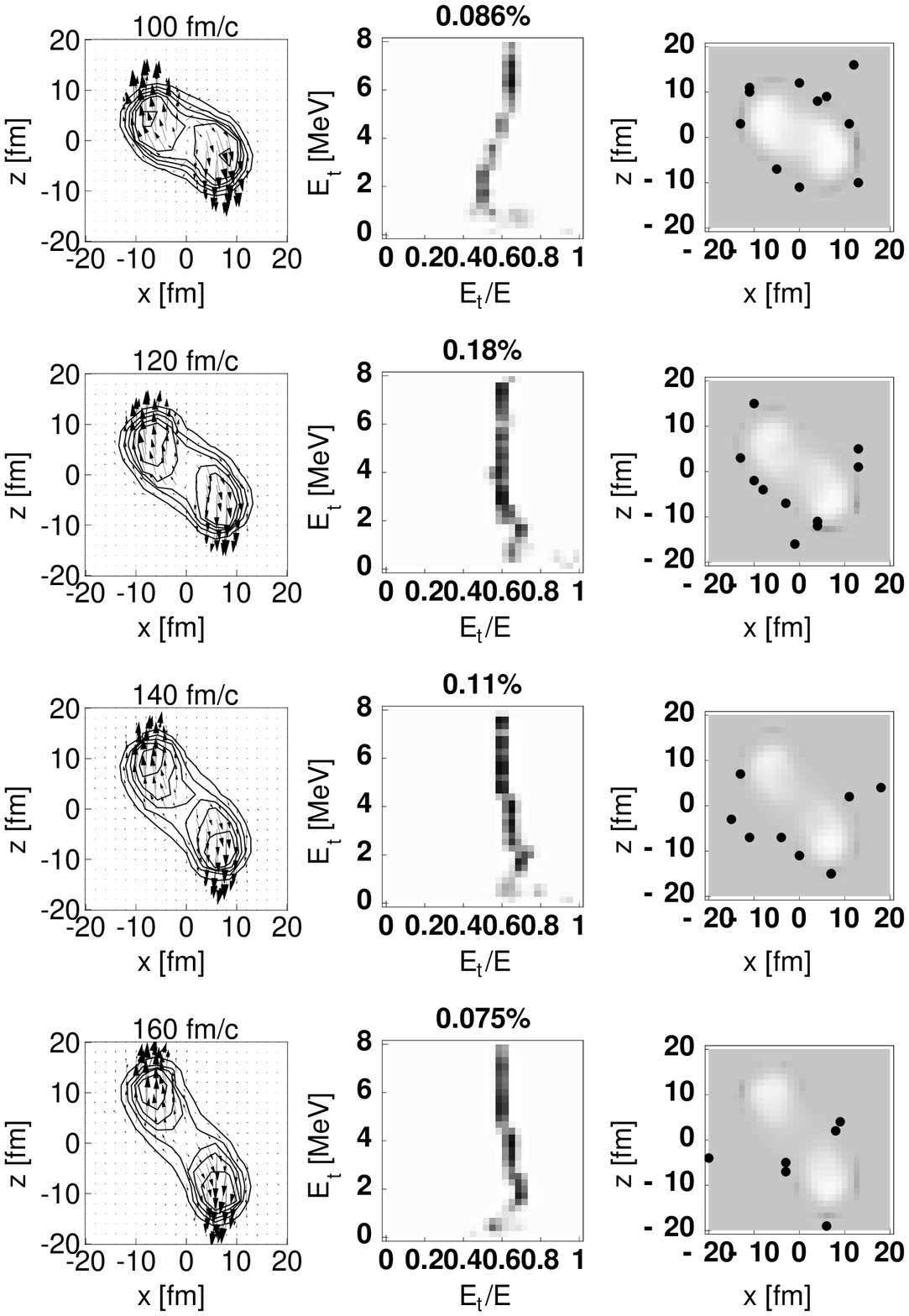,width=15cm,angle=0}}
\caption{The evolution of a $Ta+Au$ collision at $33$ MeV lab energy
  and $8$ fm
  impact parameter. The left panel gives the spatial density
  contour-plot in the $(x,0,z)$ plane and the local currents as
  arrows. 
The middle panel 
shows the transversal energy $E_t$ distribution versus the ratio of
transversal to total energy $E_t/E$. The amount of emitted particles 
due to the squeezed
mode which is located in the right lower corner is given as plot
label. 
The right panel represents the contour-plots of
the total energy of particles. The dark areas gives the positive total
energy indicating areas where particle can escape and the lighter colour
scales the deepness of negative total energy indicating bound states.
The black dots are the position of test particles in the $(x,0,z)$
plane which contribute to the new squeezing mode. }\label{1}
\end{figure}

In order to verify the existence of such a mode we solved the nonlocal kinetic equation (nonlocal
BUU) \cite{SLM96,MLSCN98,MT00}
 which leads
to figure \ref{1}. We have used a soft parameterisation of the
mean-field and realistic nuclear potentials, for details see \cite{MLSK98}. 
In the second panel 
we plotted the total kinetic energy of
free particles including their Fermi motion. Calculating the current
one could get rid of the latter motion \cite{MT00} in order to come
nearer to the experiments which leads of course to a main almost linear
correlation starting from zero. An alternative way would be to use coalescence
models which we want not to use here in order to maintain theoretical consistency. 
The particles are
considered as free if their kinetic energy overcomes the binding mean
field energy. 

We see a separation of the energy distribution
of emitted particles into
two branches at $60$fm/c. The main line almost at $E_{trans}/E=2/3$
are the emitted particles due to thermal emission and
multifragmentation. Besides this line we recognise some events in the
right lower corner of this plot. These $0.2\%$ of emitted particles are due to
the squeezing mode since they have distinctly smaller energy than the
rest of emitted particles and are clearly transversal. Interestingly
this mode is seen at the time of closest approach and again later at
$120$fm/c where the neck structure appears. This can be understood
since during the time in between this mode is shadowed and screened by the
two nuclei and other emission channels.

From the right panels one sees
that these emitted particles originate really from the surface and
especially from the touching point. The number of indicated
trajectory points is not representing the amount of emitted particles
since it is only a cut in the $(x,z)$ plane. The total number is given
in the middle panel as plot-label. From this we see that the
homogeneous 
surface emitted particles at the beginning are very small and not due
to the squeezing mode. At the time of remarkable amount of squeezed
particles one sees that they come mainly from the region of
overlap. Moreover the potential surface plot shows the deepening and
squeezing of the potential at the touching point. Both observations
suggest that these emitted particles are probably due to the 
squeezing mode described above.

Let us comment that the local
Boltzmann equation (BUU) leads to an even more pronounced
effect. Since we consider the nonlocal extension of the Boltzmann
equation as more realistic we give here the smallest estimate of the
effect. Moreover, the simplified picture given above neglects
completely the rebinding mechanisms e.g. by mean fields which will
limit the ionisation. This was a sacrifice for analytical
solvability. However the realistic simulation shows that a small
percentage
of events might show this behaviour.

To summarise a new mode is predicted due to fast shrinking of
available phase space for outer particles when two nuclei collide. The
time in such collisions is too short to establish a common ionisation
threshold. Instead the energy of the level increases and ionisation
can occur. In opposition to the Mott transition, here the levels increase
which gives rise to the name ''squeezing mode''. The characteristics of such
a mode are that the emitted particles have very small total energy and
are transversal. This feature and the fact that the mode appears at
the beginning of the collision distinguishes it from  the ''towing mode''
\cite{LSC99}. In a corresponding plot we could identify such events
in nonlocal BUU simulations. In order to demonstrate this mode an
analytical solution of the
time dependent Schroedinger equation has been given.

The described mode is not restricted to nuclear collisions. Anytime
when two clusters of particles collide fast enough to prevent the
formation of a common threshold and long enough to squeeze the outer
bound states this squeezing mode should be possible to observe. 

The discussions with Jean Colin, Daniel Cussol and Jaques Normand are
gratefully acknowledged which have convinced us that this mode is possible 
to observe experimentally. Especially to Victor Dodonov, Marco Ameduri and John
Frankland we are indebted for valuable comments.

%\bibliography{kmsr,kmsr1,kmsr2,kmsr3,kmsr4,kmsr5,kmsr6,kmsr7,delay2,delay3,spin}
%\bibliographystyle{prsty}

\onecolumn

\end{document}